\documentclass[prl,twocolumn,unsortedaddress,superscriptaddress]{revtex4-2}
\usepackage[english]{babel}
\addto\captionsenglish{}
\usepackage{graphicx}

\usepackage{epsfig,color}
\usepackage{amsmath,amssymb,eufrak,calc}
\usepackage{amsthm}
\usepackage{enumerate}
\usepackage{hyperref}
\usepackage{amsxtra,amsfonts,bm,txfonts}
\usepackage{blkarray}
\usepackage{adjustbox}
\usepackage{braket}
\usepackage{adjustbox}
\usepackage{xcolor}

\begin{document}

\title{Real-space blocking of qubit variables on parallel lattice gauge theory links for quantum simulation}

\date{\today}

\author{Judy Shir}
\affiliation{Racah Institute of Physics, The Hebrew University of Jerusalem, Givat Ram, Jerusalem 91904, Israel}

\author{Erez Zohar}
\affiliation{Racah Institute of Physics, The Hebrew University of Jerusalem, Givat Ram, Jerusalem 91904, Israel}

\begin{abstract}
One of the methods proposed in the last years for studying non-perturbative gauge theory physics is quantum simulation, where lattice gauge theories are mapped onto quantum devices which can be built in the laboratory, or quantum computers. While being very promising and already showing some experimental results, these methods still face several challenges related to the interface between the technological capabilities and the demands of the simulated models; in particular, one such challenge is the need to simulate infinitely dimensional local Hilbert spaces, describing the gauge fields on the links in the case of compact Lie gauge groups, requiring some truncations and approximations which are not completely understood or controllable in the general case. This work proposes a way to obtain arbitrarily large  such local Hilbert spaces by using coarse graining of simple, low dimensional qubit systems, made of components available on most quantum simulation platforms, and thus opening the way of new types of lattice gauge theory quantum simulations.
\end{abstract}

\maketitle

Quantum simulation of lattice gauge theories (LGTs) \cite{wiese_ultracold_2013,zohar_quantum_2016,dalmonte_lattice_2016,banuls_review_2020,banuls_simulating_2020,aidelsburger_cold_2022,zohar_quantum_2022,klco_standard_2022,bauer_quantum_2023,bauer_quantum2_2023,hc23} is nowadays a lively and rapid field of research, offering quantum technology and computation based methods for dealing with challenging non-perturbative open questions in gauge theories. Such models, which describe the most fundamental interactions of nature (e.g. in the standard model of particle physics \cite{langacker_standard_2011}) involve running and strong interaction couplings, for example when one wishes to study quark confinement in quantum chromodynamics (QCD)  and similar mechanisms in other models. LGTs \cite{wilson_confinement_1974,kogut_hamiltonian_1975,kogut_introduction_1979,kogut_lattice_1983}, combined with Monte-Carlo computations, have been a prominent way of overcoming the strength of interactions, mostly for static properties (see e.g. \cite{aoki_flag_2020}). But when real-time dynamics and finitely dense fermions are considered, due to the use of Euclidean spacetime, this method becomes problematic: real-time dynamics is not directly feasible, and the sign problem arises for several finite density scenarios \cite{troyer_computational_2005}. Thus quantum simulation is a great candidate for overcoming this obstacle.

While a lot of theoretical work has been made in terms of mapping relativistic theories with local symmetries to the non-relativistic, globally invariant available simulating platforms, much work is yet to be done. The experiments carried out so far \cite{martinez_real-time_2016,bernien_probing_2017,kokail_self_2019,schweizer_floquet_2019,mil_scalable_2020,yang_observation_2020,klco_su_2020,ciavarella_trailhead_2021,semeghini_probing_2021,atas_su_2021,zhou_thermalization_2022,riechert_engineering_2022,mildenberger_probing_2022,alam_primitive_2022,gustafson_primitive_2022,ciavaraella_preparation_2022,illa_basic_2022,rahman_real_2022,rahman_self_2022,atas_simulating_2022,su_observation_2023,pardo_resource_2023,farrell_scalable_2023,zhang_observation_2023,farrell_preparations_2023,farrell_preparations2_2023,ciavarella_quantum_2023,charles_z2_2023,mueller_quantum_2023,kavaki_from_2024,farrell_quantum_2024,ciavarella_quantum_2024}, while showing remarkable results, are mostly limited to small systems and low dimensions. One particular challenge has to do with the fact that the local gauge field Hilbert spaces of LGTs with compact Lie gauge groups (such as U(1) for quantum electrodynamics (QED) or SU(3) for QCD) are of infinite dimensions, and available quantum hardware for quantum simulation and computation tasks usually involve finitely dimensional spaces. Thus, one is required to truncate these Hilbert spaces, but in a way that will preserve some of the important features of the model, and ideally will converge in a meaningful limit to the physics of the non-truncated case \cite{zohar_formulation_2015,davoudi_towards_2019,lamm_general_2019,kaplan_gauss_2018,zohar_quantum_2022}.

It is well-known that invariance under the full gauge group does not require the use of the full, non-truncated Hilbert space, as in quantum link models (QLMs) \cite{horn_finite_1981,orland_lattice_1990,brower_qcd_1999,wiese_from_2022}. Such models are interesting candidates for quantum simulation on their own, showing very rich and interesting non-perturbative gauge theory physics (see, e.g., the early quantum simulation proposals \cite{banerjee_atomic_2012,banerjee_atomic_2013,tagliacozzo_simulation_2012} and plenty of other works cited in the review papers mentioned above). In order to obtain non-truncated Hilbert space physics with such models, as that of the continuum gauge theories, one needs to introduce an extra dimension as explained in \cite{brower_qcd_1999}.
The type of Hilbert space truncation done in QLMs is called representation (or irrep) basis truncation \cite{zohar_formulation_2015,davoudi_towards_2019,zohar_quantum_2022}, because it is based on using a finite subset of a basis spanned by states labeled by irreducible representations of the gauge group (another truncation method is in the dual, group element basis \cite{zohar_digital_2017,lamm_general_2019,zohar_quantum_2022}). Another option for obtaining the full Hilbert space limit, besides integrating over an extra dimensions, is including more and more representations and raising the truncation cutoff \cite{zohar_simulating_2012,zohar_formulation_2015}. Then, no extra dimension is needed, but one does have to experimentally implement larger Hilbert spaces.

This problem, however, is not new or unique to the world of quantum simulation and the Hamiltonian formulation. Similar issues were met by classical calculations when dealing with errors and their scaling with the lattice spacing, leading to the development of the improved Symanzik actions \cite{symanzik_continuum_1983}. These ideas have led to the recent development of improved lattice Hamiltonians, aimed at dealing with the truncation problem in lattice gauge theory quantum simulations \cite{carena_improved_2022,ciavarella_quantum_2023}.
Another possible approach is considering real-space renormalization, coarse graining or blocking of local degrees of freedom residing on smaller local Hilbert spaces. In the classical approach of actions and path integrals, such ideas have been introduced and studied in the past - see, e.g. works on perfect actions and fixed-point actions \cite{bietenholz_fixed_1994,bietenholz_perfect_1996,bietenholz_perfect2_1996,bietenholz_progress_1997,bietenholz_perfect_1997,bietenholz_perfect_1998,bietenholz_improved_1998,bietenholz_preconditioning_1999}, Migdal-Kadanoff and real space renormalization group \cite{migdal_phase_1975,kadanoff_notes_1976,fradkin_real_1979}, tensor renormalization group \cite{levin_tensor_2007,liu_exact_2013,meurice_tensor_2022} and more.

In this work we suggest another way to obtain the full Hilbert space, from a truncated lattice model, or at least to increase the truncation. For that, we use the fact that several lattice models may give rise to the same continuum limit, and thus introduce a model where the local Hilbert spaces are finite and very small (two level systems - spin $1/2$s or qubits, as desired for many quantum simulation implementations), but the interactions and couplings are tailored in a proper way which allows to effectively block, or coarse grain them, in a way which gives rise to arbitrarily large local Hilbert spaces. This is based on the fact that, as we shall briefly show, since the truncated quantity is the electric field, or flux, and the integral Gauss law is flux-additive. We will focus here on the simplest compact Lie group case, of compact QED (cQED) \cite{kogut_introduction_1979}, where the gauge group is U(1).

\emph{Compact QED.} Let us briefly review the basics of Hamiltonian compact QED \cite{kogut_introduction_1979}. We consider a spatial $d$ dimensional lattice, whose sites are labeled by integer vectors $\mathbf{x}\in\mathbb{Z}^d$. The matter degrees of freedom reside on the sites; for simplicity, we make them staggered \cite{susskind_lattice_1977}, and therefore each site can host at most one fermion created by $\psi^{\dagger}\left(\mathbf{x}\right)$. The gauge fields reside on the links; each link of the lattice, labeled by a pair of a site $\mathbf{x}$ and a direction $i=1...d$, for the link's starting point and direction respectively, hosts the Hilbert space of a particle on a ring, where a phase (or an angle) operator $\phi\left(\mathbf{\mathbf{x}},i\right)$ and its conjugate angular momentum operator $E\left(\mathbf{x},i\right)$ are defined. The spectrum of $\phi$ takes the values $\left[0,2\pi\right)$, while $E$ has an integer, non-bounded spectrum, as a $U(1)$ angular momentum operator. $\phi$ plays the role of the (compact) vector potential and $E$ is the electric field. Since they are canonically conjugate, $\left[\phi,E\right]=i$, the unitary operator $U=e^{i\phi}$ serves as a raising operator for the electric field on the link - $\left[E,U\right]=U$; that is, if we denote the electric field eigenstates by $\left\{\left|m\right\rangle\right\}_{m=-\infty}^{\infty}$, such that $E\left|m\right\rangle=\left|m\right\rangle$, we have
$	U\left|m\right\rangle=\left|m+1\right\rangle$.
The Hamiltonian of the model, $H=H_{\text{m}}+H_{\text{E}}+H_{\text{B}}+H_{\text{int}}$ includes four parts. The \emph{mass part}, of the fermions alone, takes in the staggered case the form $H_{\text{m}}=m\underset{\mathbf{x}}{\sum}\left(-1\right)^{x_1+...+...x_d}\psi^{\dagger}\left(\mathbf{x}\right)\psi\left(\mathbf{x}\right)$,
 where $m$ is the mass. The \emph{electric part} which measures the electric energy on the lattice's links takes the form $H_{\text{E}}=\frac{g^{2}}{2}\underset{\mathbf{x},i}{\sum} E^2\left(\mathbf{x},i\right)$, where $g$ is the coupling constant. The \emph{magnetic part} which corresponds, in the classical continuum limit, to the well-known magnetic energy of QED, involves four body interactions on plaquettes (plaquette terms),
 $H_{\text{B}}=-\frac{1}{g^{2}}\underset{\mathbf{x},i<j}{\sum}\cos\left(
 \phi\left(\mathbf{x},i\right) +
 \phi\left(\mathbf{x}+\hat{\mathbf{e}}_i,j\right) -
 \phi\left(\mathbf{x}+\hat{\mathbf{e}}_j,i\right) -
 \phi\left(\mathbf{x},j\right) 
 \right)$, where $\hat{\mathbf{e}}_i$ is a unit (lattice) vector in direction $i$. Finally, the matter and gauge fields are minimally coupled through the \emph{interaction Hamiltonian},
 $H_{\text{int}}=\underset{\mathbf{x},i}{\sum}\left(t_i\left(\mathbf{x}\right)\psi^{\dagger}\left(\mathbf{x}\right)
 U\left(\mathbf{x},i\right)
 \psi\left(\mathbf{x}+\hat{\mathbf{e}}_i\right) + \text{H.c.} \right)$,
 where the hopping amplitudes $t_i\left(\mathbf{x}\right)$ may be chosen in a way which reproduces a relativistic continuum (Dirac) limit \cite{susskind_lattice_1977}.
 
 Gauge transformations are local unitary operations generated by the \emph{Gauss law operators}, 
 $G\left(\mathbf{x}\right)=\underset{i}{\sum}\left(
 E\left(\mathbf{x},i\right)-
 E\left(\mathbf{x}-\hat{\mathbf{e}}_i\right)\right)-Q\left(\mathbf{x}\right)$, which is the difference between the lattice divergence of the electric fields at the site $\mathbf{x}$ and the local fermionic charge $Q\left(\mathbf{x}\right)$. In the staggered case, for example, 
 $Q\left(\mathbf{x}\right) = \psi^{\dagger}\left(\mathbf{x}\right)\psi\left(\mathbf{x}\right)
 - \frac{1}{2}\left(1-\left(-1\right)^{x_1+...+x_d}\right)$ \cite{susskind_lattice_1977}. These operators all commute with the Hamiltonian - $\left[H,G\left(\mathbf{x}\right)\right]=0$ for all $\mathbf{x}$, and thus they generate a local symmetry. \emph{Physical states} are eigenstates of these, which can be simultaneously diagonalized with the Hamiltonian. The eigenvalues are usually called static charges; we will focus here on the case where no static charges exist (merely for simplicity - they can be introduced in a straightforward fashion), and thus the physical states are those satisfying
 \begin{equation}
 	G\left(\mathbf{x}\right)\left|\psi\right\rangle = 0,\quad\forall \mathbf{x}.
 \end{equation}
These constraint are the \emph{Gauss laws}, and the states which obey them are \emph{gauge invariant}.

\emph{Truncating the Electric Field.} A common problem in the design of quantum simulators of various physical models is the need to approximate and express infinitely dimensional local Hilbert spaces in terms of finite ones. For example, if only qubit or qudit degrees of freedom are available, or in rather more analog schemes if a finite set of states is available, e.g. when atoms or ions are used, only with a finite number of accessible internal levels which can be addressed and manipulated by the experimental device. In particular, since the Hilbert space on a link of a lattice gauge theory with a compact Lie gauge group has an infinite dimension, we run into this problem.

As explained in the introduction, one of the ways out of it has been the truncation of the electric field, such that $E$ can only take values between $-\ell$ and $\ell$. Then, we can think about the possible electric field state on a link as the SU(2) multiplet $\left\{\left|\ell m\right\rangle\right\}_{m=-\ell}^{\ell}$, where $\ell$ is fixed (either integer or half integer); the electric field is then well represented by the $L_z$ operator, satisfying $L_z\left|\ell m\right\rangle=m\left|\ell m\right\rangle$. Its raising and lowering is done by the operators $L_{\pm}$ for which $\left[L_z,L_{\pm}\right]=\pm L_{\pm}$. Therefore, if one performs the replacements $E \rightarrow L_z$ and $U \rightarrow L_{+}$ in the Hamiltonian and Gauss laws, we get a lattice gauge theory with the same, full $U(1)$ gauge invariance, but finite gauge field Hilbert spaces. In such a truncation,  the operator $\phi$ and its eigenstates, forming the group element basis are lost and the raising / lowering operators, now acting on a finite ladder of electric fields, are no longer unitary (but one can preserves these instead, when the truncation is made in the other basis \cite{zohar_digital_2017,lamm_general_2019,zohar_quantum_2022}).

Electric field truncation is considered to be a good enough approximation within a confining phase, where the electric field is mostly bounded and concentrated, but not in other phases. This can be seen as a weak point when one wishes to study all various phases, or when the phase transition point or the required minimal $\ell$ is not well known. On the other hand it is clear that as $\ell$ grows towards infinity, the original physics of the non-truncated model is better approximated; for a fixed integer $\ell$ truncation define the operators
$V = L_{+} / \sqrt{\ell\left(\ell+1\right)}$. Note that for $\left|m\right| \ll \ell$, we have
\begin{equation}
V \left|\ell m\right\rangle = \sqrt{1-\frac{m\left(m+1\right)}{\ell\left(\ell+1\right)}} \left|\ell ,m+1\right\rangle
\underset{\left|m\right| \ll \ell}{\approx} \left|\ell ,m+1\right\rangle,
\end{equation}
approximating the unitary action of $U=e^{i\phi}$ \cite{zohar_simulating_2012}. One can therefore choose to replace the $U$ operators in the original non-truncated model by the $V$ ones. This was shown to give excellent results of the non-truncated with finite $\ell$ values in \cite{Kasper_Implementing_2017}. But in many cases the simulating platform does not allow for very high $\ell$ values, as was possible, for example, in the case of \cite{Kasper_Implementing_2017} and the respective experiment \cite{mil_scalable_2020} where the gauge field Hilbert space is obtained by using the Schwinger algebra of two bosonic species, with a fixed total number, trapped on a link. Quite often, it is easier to implement interactions with the smallest Hilbert space dimension on the link, for example that of a single qubit. There, the electric field is within the $\ell=\frac{1}{2}$ representation - not even an integer one. While it gives rise to the right symmetry and can be implemented (e.g. the first QLM realization \cite{banerjee_atomic_2012} and many following works), it is not necessarily straight-forward to construct from it a Hilbert space with an arbitrarily large integer $\ell$. This is what we shall now proceed to do.

\emph{The building blocks - ladder physics.}
Consider an open chain of $N=2\ell$ spins, where $\ell$ is an integer. We label them by $n=1,...,N$, and denote by $\mathbf{S}^{(n)}=\frac{1}{2}\mathbf{\sigma}$ their spin operators ($\mathbf{\sigma}$ is the vector of Pauli matrices). 
 introduce the simple Hamiltonian
\begin{equation}
	h_{\text{L}}=-2\lambda\underset{n=1}{\overset{N-1}{\sum}}\mathbf{S}^{(n)}\cdot\mathbf{S}^{(n+1)}=
	-\lambda\underset{n=1}{\overset{N-1}{\sum}}\left(\mathbf{S}^{(n)}+\mathbf{S}^{(n+1)}\right)^2 + \text{Const.}
\end{equation}
Since all the operators $\left(\mathbf{S}^{(n)}+\mathbf{S}^{(n+1)}\right)^2$, the eigenstates of $h_{\text{L}}$ are states where the state of every two neighboring spins is of a fixed total spin (zero or one, from the addition of two $1/2$ spins). We focus on $\lambda>0$, and thus in a ground state of $h_{\text{L}}$, every two neighboring links will be in a spin one state. Therefore, any ground state of $h_{\text{L}}$ will be symmetric under the exchange of two neighboring spins, and hence under the exchange of any two spins (by composition of nearest neighbor exchanges - see the appendix for details). This complete exchange symmetry implies that a ground state of $h_{\text{L}}$ will be in the highest possible spin multiplet - in this case, $\ell$. Being of the highest spin, this multiplet will have multiplicity $1$, and we conclude that $h_{\text{L}}$ has exactly $2\ell+1$ degenerate ground states,  $\left\{\left|\ell m\right\rangle\right\}_{m=-\ell}^{\ell}$. We denote the projector onto this ground space by
\begin{equation}
	P_{\ell} = \overset{\ell}{\underset{m=-\ell}{\sum}}\left|\ell m\right\rangle\left\langle\ell m\right|.
\end{equation}

\begin{figure}[t!]
	\centering
	\includegraphics[width=\columnwidth]{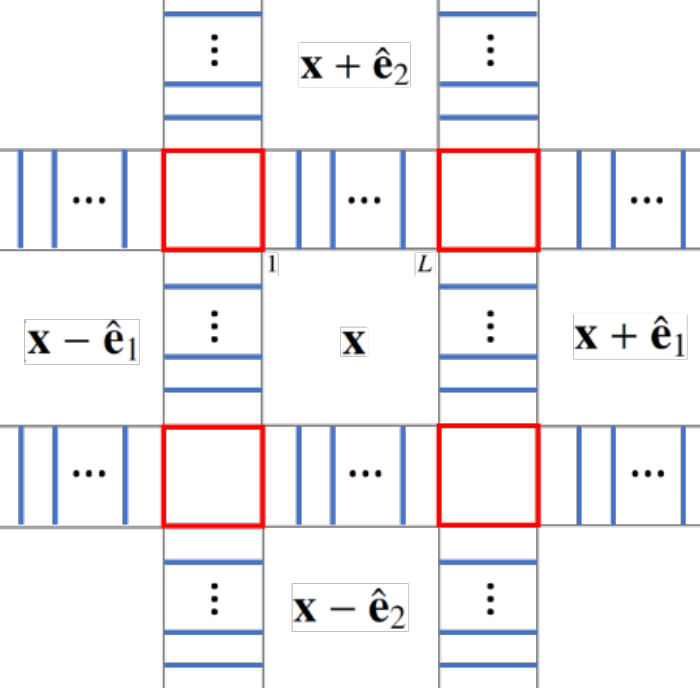}
	\caption{The primitive system, to be coarsed grained. The gray squares are the blocks, corresponding to single sites. Between them, to each direction, ladders of $N$ parallel links stretch, labeled from $1$ to $N$ - from left to right and from bottom to top. A single spin, or qubit, occupies each of them. They are marked by blue and red; the red ones form the primitive plaquette interactions of $H^{(0)}_{\text{B}}$. The blue ones only participate in the ladder terms, $H_{\text{L}} +  H^{(0)}_{\text{E}}$.
	When matter is added, primitive modes $\chi$ are placed on the boundary of the blocks. 
	$H^{(0)}_{\text{b}}$ involves modes within a block - that is, around its boundary. $H^{(0)}_{\text{int}}$ couples primitive fermionic modes on two neighboring blocks, across a ladder, including the ladder's links.
	}
	\label{figfig}
\end{figure}

\emph{The primitive model and its coarse graining.} We now consider, for simplicity, a two dimensional square lattice, and label its sites by $\mathbf{x}\in\mathbb{Z}^2$. We think of these sites as blocks of a more primitive, intrinsic lattice, with $N \times N$ sites. Thus, between two neighboring blocks a ladder of $N = 2\ell$ links is stretched, and we place a single spin $1/2$ on each link (see Fig. \ref{figfig}).
Our goal is to obtain a blocked, coarse grained lattice, where all the parallel links in a ladder are grouped together to a single link with truncation $\ell$. On such an effective link, emanating from the block $\mathbf{x}$ to its nearest neighbor $\mathbf{x} + \hat{\mathbf{e}}_i$, we introduce the total spin operators, $\mathbf{L}\left(\mathbf{x},i\right) = \overset{N}{\underset{n=1}{\sum}}\mathbf{S}^{(n)}\left(\mathbf{x},i\right)$, where $n$ denotes the spin's position along the ladder.

Consider the Hamiltonian
\begin{equation}
H^{(0)} = H_{\text{L}} +  H^{(0)}_{\text{E}} +  H^{(0)}_{\text{B}},
\end{equation}
where $H_{\text{L}}=\underset{\mathbf{x},i}{\sum}h_L\left(\mathbf{x},i\right)$ - a sum of the ladder operators introduced above on all the ladders; 
\begin{equation}
	H^{(0)}_{\text{E}} =  g^2\ell\underset{\mathbf{x},i}{\sum}\overset{N-1}{\underset{n=1}{\sum}}
	S_z^{(n)}\left(\mathbf{x},i\right)S_z^{(n+1)}\left(\mathbf{x},i\right)
\end{equation} is a sum of $zz$ nearest-neighbor spin interactions within the ladders; and
\begin{widetext}
	\begin{equation}
H^{(0)}_{\text{B}} = -\frac{1}{32g^2}\left(\frac{\ell}{\ell+1}\right)^2 
\underset{\mathbf{x}}{\sum}\left(
S_+^{(L)}\left(\mathbf{x},1\right)
S_+^{(1)}\left(\mathbf{x}+\hat{\mathbf{e}}_1,2\right)
S_-^{(1)}\left(\mathbf{x}+\hat{\mathbf{e}}_2,1\right)
S_-^{(L)}\left(\mathbf{x},2\right)
+\text{H.c.}\right)
\end{equation}
\end{widetext}
is a plaquette interaction of the first/last links of the ladders, closing squares at the corners of the blocks (see Fig. \ref{figfig}).

All these interactions are quite simple and straightforward to implement with a variety of quantum hardware, e.g. atomic systems, nano-photonical devices, superconducting qubits and more, in analog or digital manners. We do not specify here the prescriptions for implementing such terms experimentally, in order to emphasize the generality of the scheme and not imply that it must be narrowed down to a particular hardware or simulation approach. Nevertheless, we stress that the above qubit interaction terms required for the quantum simulation of the pure gauge part of our Hamiltonian are fairly simple.

Note that the Hamiltonian $H^{(0)}$ is invariant under gauge transformations generated by 
$G\left(\mathbf{x}\right) = L_z\left(\mathbf{x},1\right) + L_z\left(\mathbf{x},2\right) - L_z\left(\mathbf{x}-\hat{\mathbf{e}}_1,1\right) - L_z\left(\mathbf{x}-\hat{\mathbf{e}}_2,2\right)$ - pure gauge transformations generated by the \emph{blocked} electric fields. We will now proceed to obtain an effective Hamiltonian which is written in terms of the blocked or coarse grained Hilbert spaces, where the ladders become the links; but since the primitive Hamiltonian before blocking is already gauge invariant under the blocked generators, it implies that the symmetry will be exact.

To obtain the coarse grained Hamiltonian, we choose $\lambda$ such that $\lambda \gg g^2\ell$ as well as $\lambda \gg \frac{8}{g^2}\left(\frac{\ell+1}{\ell}\right)^2 $. Being the largest energy scale of the problem, it is reasonable to develop an effective Hamiltonian with respect to the ground sector of $H_{\text{L}}$ \cite{soliverez_effective_1969,cohen-tannoudji_atom-photon_1992}. We will do this to the lowest order in perturbation theory, simply by projecting all the terms of $H^{(0)}$ onto the relevant subspace, with the projector
$\mathcal{P}_{\ell} = \underset{\mathbf{x},i}{\prod}P_{\ell}\left(\mathbf{x},i\right)$. The effective Hamiltonian will take the form $H = \mathcal{P}_{\ell} H^{(0)} \mathcal{P}_{\ell}$.

Projecting $H_{\text{L}}$ which gives us the constraint will give rise to a constant, which we will ignore. $	H^{(0)}_{\text{E}}$ is a sum of decoupled ladder terms, which we can treat separately. Note that since the states with total spin $\ell$ are symmetric, we have that $P_{\ell} S^{(n)}_z S^{(k)}_z P_{\ell} = P_{\ell} S^{(n')}_z S^{(k')}_z P_{\ell}$, for any $n \neq k$, $n' \neq k'$. Therefore,
$\mathcal{P}_{\ell} H^{(0)}_{\text{E}} \mathcal{P}_{\ell} =  g^2\ell\underset{\mathbf{x},i}{\sum}\overset{N-1}{\underset{n=1}{\sum}}
\mathcal{P}_{\ell}S_z^{(1)}\left(\mathbf{x},i\right)S_z^{(2)}\left(\mathbf{x},i\right) \mathcal{P}_{\ell}
= \frac{ g^2}{2}N\left(N-1\right)\underset{\mathbf{x},i}{\sum}
\mathcal{P}_{\ell}S_z^{(1)}\left(\mathbf{x},i\right)S_z^{(2)}\left(\mathbf{x},i\right) \mathcal{P}_{\ell}
$. On the other hand, note that since
$L_z^2\left(\mathbf{x},i\right) = 2\overset{N}{\underset{n=1}{\sum}}
\overset{N}{\underset{n'=n+1}{\sum}}S_z^{(n)}\left(\mathbf{x},i\right)S_z^{(n')}\left(\mathbf{x},i\right) + \text{Const.}$, we get that 
$\mathcal{P}_{\ell}L_z^2\left(\mathbf{x},i\right)\mathcal{P}_{\ell}
=2\frac{N\left(N-1\right)}{2}\mathcal{P}_{\ell}S_z^{(1)}\left(\mathbf{x},i\right)S_z^{(2)}\left(\mathbf{x},i\right) \mathcal{P}_{\ell} + \text{Const.}
$,
or simply:
\begin{equation}
	\mathcal{P}_{\ell} H^{(0)}_{\text{E}} \mathcal{P}_{\ell} = \frac{g^2}{2}\underset{\mathbf{x},i}{\sum}L_z^{2}\left(\mathbf{x},i\right) + \text{Const.} = H_{\text{E}}  + \text{Const.}
\end{equation}
 - the desired, post-blocking electric Hamiltonian (where we used the fact that $L_z$ and its powers do not mix sectors of $H_{\text{L}}$). Note that since this part of the Hamiltonian commutes with the constraints, this is the full expansion in this term.
 
 Finally, let us consider the magnetic part. Note that for any $n,n'$, $P_{\ell} \mathbf{S}^{(n)} P_{\ell} = P_{\ell} \mathbf{S}^{(n')} P_{\ell}$, thanks to the full exchange symmetry. Therefore, $P_{\ell} \mathbf{L} P_{\ell} = L P_{\ell} \mathbf{S} P_{\ell}$ and we get $P_{\ell} S^{(n)}_{\pm} P_{\ell} = \frac{2}{\ell} P_{\ell}L_{\pm}P_{\ell}$, and we get that the plaquette terms become
\begin{widetext}
	\begin{equation}
		\mathcal{P}_{\ell} H^{(0)}_{\text{B}} \mathcal{P}_{\ell}= -\frac{1}{2g^2}
		\underset{\mathbf{x}}{\sum}\left(
		V\left(\mathbf{x},1\right)
		V\left(\mathbf{x}+\hat{\mathbf{e}}_1,2\right)
		V^{\dagger}\left(\mathbf{x}+\hat{\mathbf{e}}_2,1\right)
		V^{\dagger}\left(\mathbf{x},2\right)
		+\text{H.c.}\right) = H_{\text{B}}
	\end{equation}
\end{widetext}
- the desired, post-blocking magnetic Hamiltonian.

\emph{Inclusion of matter.} To add matter, we introduce primitive fermionic modes at the beginning and end of each primitive ladder link - that is, on the sites which are on the block's boundary (see Fig. \ref{figfig}). We create these fermions using operators of the form $\chi^{\dagger}$, one on each primitive boundary site, altogether $4\left(N-1\right)$ modes per block. We label them by $a=1,...,4\left(N-1\right)$ - some periodic coordinate around the block's boundary, starting, for example, from the lowest right corner. The fermionic modes are described by the additional Hamiltonian
$H^{(0)}_{\text{F}} = H^{(0)}_{\text{b}} + H^{(0)}_{\text{int}}$, where $H^{(0)}_{\text{b}}$ describes the modes within a block and $H^{(0)}_{\text{int}}$ the interaction with the neighboring blocks through the intermediate ladders.
More specifically,
$H^{(0)}_{\text{b}} = \underset{\mathbf{x}}{\sum}\left(m\left(-1\right)^{x_1+x_2}\underset{a}{\sum}\chi^{\dagger}_a\left(\mathbf{x}\right)
\chi_a\left(\mathbf{x}\right)+\Delta\underset{a}{\sum}\left(\chi^{\dagger}_a\left(\mathbf{x}\right)\chi_{a+1}\left(\mathbf{x}\right)+\text{H.c.}\right)\right)$ (where $a=4L-3$ corresponds periodically to $a=1$) and 
$H^{(0)}_{\text{int}}
=
\frac{8\ell-4}{\sqrt{\ell\left(\ell+1\right)}}
\underset{\mathbf{x}}{\sum}t_1\left(\mathbf{x}\right)\overset{L}{\underset{a=1}{\sum}}\chi_a^{\dagger}\left(\mathbf{x}\right)S^{(a)}_+\left(\mathbf{x},1\right)\chi_{3N-1-a}\left(\mathbf{x}+\hat{\mathbf{e}}_1\right)
+
\frac{8\ell-4}{\sqrt{\ell\left(\ell+1\right)}}
\underset{\mathbf{x}}{\sum}
t_2\left(\mathbf{x}\right)\overset{2N-1}{\underset{a=N}{\sum}}\chi_a^{\dagger}\left(\mathbf{x}\right)S^{(a-N+1)}_+\left(\mathbf{x},2\right)\chi_{4N-3-a}\left(\mathbf{x}+\hat{\mathbf{e}}_2\right)
+ \text{H.c.}
$
Here again we see terms that are easy to implement in a variety of simulating systems, including simple link interactions whose preparation was suggested in a variety of quantum simulation proposals for quantum link models, which can be found, for example, in most of the review articles cited in the introduction.

Assuming that $\Delta \gg m, \frac{8\ell-4}{\sqrt{\ell\left(\ell+1\right)}}$ and that $\lambda \gg m,\frac{8\ell-4}{\sqrt{\ell\left(\ell+1\right)}}$ we can carry on the effective analysis, and project onto the ground state of the $\Delta$ dependent term. We further assume that each block may contain up to a single fermion. A fourier transform in the quasi-momentum conjugate to $a$ will leave us effectively only with the zero momentum mode, which we denote as the one created by $\psi^{\dagger}$. The mass term commutes with the constraints. In the interaction terms, the projection of each $\chi^{\dagger}$ or $\chi$ will give rise to $\frac{1}{\sqrt{4\left(L-1\right)}}\psi^{\dagger}$ or $\frac{1}{\sqrt{4\left(N-1\right)}}\psi$; each $S^{(i)}_{\pm}$ will be projected as before to $\frac{1}{N}N_{\pm}$, and each of the $N$ link interactions of each ladder will be projected to the same thing. Therefore, the overall factor will be of $\frac{N}{4N\left(N-1\right)} = \frac{1}{8\ell-4}$. Altogether we get that the effective Hamiltonian involving the fermions is
	\begin{equation}
		H_{\text{F}} = H_{\text{m}}+H_{\text{int}}
	\end{equation}
as desired for the post-blocked system, where $H_{\text{int}}$ contains $V$ instead of $U$.

To conclude the matter inclusion part, let us briefly discuss whether the choice of staggered fermions, made here for simplicity, could be generalized to other prescriptions of lattice fermions. Staggered fermions form a convenient choice here, for the simple reason that they are described by single spin component per site. Clearly, adding a spin indices - more spin components per site, as in the cases of naive and Wilson fermions, does not change the procedure; one needs to add more $\chi$ modes, one per spin component, and introduce identical copies of $H^{(0)}_{\text{b}}$ which do not mix them. Proper spin component mixing in $H^{(0)}_{\text{int}}$ will give rise to the right effective Hamiltonian. These are the types of lattice fermions mostly discussed in the context of quantum simulation (see, e.g. \cite{gustafson_improved_2024}), mostly thanks to the fact that they do not include further theoretical requirements which are hard to implement on experimental hardware. Further cases, such as Ginsparg-Wilson or overlap fermions, to which a Hamiltonian formulation was recently introduced \cite{clancy_toward_2023}, will require further study, and the generalization to them is not straight-forward or guaranteed.

\emph{Higher dimensions.} The generalization to higher space dimension is straight-forward. For example, in three space dimensions, the blocks are cubic instead of square, fermionic modes will be attached to their boundaries again, and the ladders are now two dimensional grids of parallel links. All of the above can be generalized in a direct fashion.

\emph{Summary.} To conclude, we see that the quantum simulation of a $U(1)$ lattice gauge theory, with a fairly general electric field truncation, can be obtained by properly choosing the energy scales and strength of interaction in a fairly simple Hamiltonian, containing standard couplings and interactions between qubits (in the pure gauge case) and fermions (if matter is introduced). One needs to prepare the states on the ladders initially in the right multiplet, but this can be done by first setting all the spins to point up or down, giving us for sure the states $\left|\ell,\pm \ell\right\rangle$. From them it is possible to obtain any other element of the right multiplet without leaving it, by acting with the ladder operators. As for measurements, one should note that matrix elements and expectation values of the blocked operators  correspond, in our sector, to such of the qubit operators, and thus simple operations can be used here too. High dimension generalizations are straightforward, and hence one could think of designing particular experiments tailored to specific systems implementing the proposed approach.

Non-trivial future extension which will require serious considerations, and are not even guaranteed to be possible, involve generalizations of the method to non-Abelian groups, as well as the discussion of the continuum limits.

\emph{Acknowledgments.} E.Z. acknowledges  the support of the Israel Science Foundation
	(grant No. 523/20). 

\section*{Appendix: addition of an even number of spins}

In this appendix we review some basic properties used throughout this work on the addition of many spin-$1/2$ particles. These results are not new and may either be found in quantum mechanics textbooks, or be easily proven based on them; yet, we include them here for completeness and clarity.

We consider an even number, $N=2\ell$, of spins, where $\ell$ is an integer. Each of them is described by a vector of spin operators, $\mathbf{S}^{(n)} = \left(S^{(n)}_x,S^{(n)}_y,S^{(n)}_z\right)$, related to the Pauli matrices through
$S^{(n)}_a = \sigma^{(n)}_a /2$. When we add them, we define the total spin operators
\begin{equation}
	\mathbf{L} = \overset{N}{\underset{n=1}{\sum}}\mathbf{S}^{(n)}.
\end{equation}
In such an addition of spins, we get multiplets of any integer spin from $0$ to $\ell$. The spins $0,...,\ell-1$ come with multiplicities larger than 1, which may be computed, but the highest spin multiplet, $\ell$, has no multiplicity and appears once. We are interested in this highest multiplet - states of the form $\left|\ell m\right\rangle$ with $m=-\ell,...,\ell$, and would like to prove a few statements of them which are used throughout this work.

\emph{Statement 1: all the states $\left\{\left|\ell m\right\rangle\right\}_{m=-\ell}^{\ell}$ of the highest spin multiplet are completely symmetric under the exchange of any two spins. And vice versa: if a state is symmetric under the exchange of any two spins, the total spin is $\ell$.}

We begin with the first direction. Let $X_{nn'}$ be an operator which exchanges spins $n$ and $n'$. Obviously, it is unitary and hermitian (since it squares to itself). 
Consider the state $\left|\ell \ell\right\rangle$. For any choice of two spins $n \neq n'$, we can decompose it, using Clebsch-Gordan coefficients, to the combined spin of these two and the rest:
\begin{equation}
	\left|\ell \ell\right\rangle =
	\underset{J,M,J',M'}{\sum}\left\langle J M , J'M' |\ell \ell \right\rangle \left|JM\right\rangle_{nn'} \otimes \left|J'M'\right\rangle_{\text{rest}}.
\end{equation}
Since $\left|JM\right\rangle_{ij}$ is obtained from the addition of two spins, $J=0,1$. Since $\left|J'M'\right\rangle_{\text{rest}}$ is obtained from the addition of $2\left(\ell-1\right)$ spins, $J'=0,...,\ell-1$.
However, since the Clebsch-Gordan coefficients satisfy
\begin{equation}
	\left\langle J M , J'M' |\ell \ell \right\rangle \propto\delta_{M+M',\ell},
\end{equation}
we get that $J=1$: otherwise, for $J=0$, since the maximal value of $M'$ is $\ell-1$, the Clebsch-Gordan coefficients are all zero. The states $\left|1M\right\rangle_{nn'}$ are completely symmetric under the exchange of $n,n'$, that is,
\begin{equation}
	X_{nn'}\left|1M\right\rangle_{nn'}=\left|1M\right\rangle_{nn'}  ,\quad\forall n \neq n' .
\end{equation}
Since we can decompose $\left|\ell \ell\right\rangle$ that way for any $n\neq n'$ pair, we conclude that it is symmetric under the exchange of any two spins,
\begin{equation}
	X_{nn'}\left|\ell \ell\right\rangle=\left|\ell \ell\right\rangle  ,\quad\forall n \neq n' .
\end{equation}

It is easy to see that $X_{nn'}$ commutes with the total spin generators, that is,
\begin{equation}
\left[\mathbf{L},X_{nn'}\right]=0 ,\quad\forall n \neq n' ,
\end{equation}
and hence, also
\begin{equation}
	\left[L_{\pm},X_{nn'}\right]=0 ,\quad\forall n \neq n' .
\end{equation}
We know that for any $m=-\ell,...,\ell$,
\begin{equation}
	\left|\ell m\right\rangle \propto L_-^{\ell-m}\left|\ell \ell\right\rangle,
\end{equation}
and thus we conclude that
\begin{equation}
	X_{nn'}\left|\ell m\right\rangle=\left|\ell m\right\rangle  ,\quad\forall i \neq j , m=-\ell,...,\ell,
\end{equation}
which proves the first direction. 

To prove the other direction, let $\left|\psi\right\rangle$ be a state such that
\begin{equation}
	X_{nn'}\left|\psi\right\rangle = \left|\psi\right\rangle,\quad\forall n \neq n' .
\end{equation}
We would like to prove that $\mathbf{L}^2 \left|\psi\right\rangle = \ell\left(\ell+1\right)\left|\psi\right\rangle$.

To see that, write
\begin{equation}
	\mathbf{L}^2 = \left(\overset{2\ell}{\underset{n=1}{\sum}}\mathbf{S}^{(n)}\right)^2 
	= \overset{2\ell}{\underset{n=1}{\sum}}\mathbf{S}^{(n)2}
	+ 2\overset{2\ell}{\underset{n=1}{\sum}}\overset{2\ell}{\underset{n'=n+1}{\sum}}
	\mathbf{S}^{(n)} \cdot \mathbf{S}^{(n')}.
\end{equation}
Since all the initial spins are $1/2$, we have $\mathbf{S}^{(n)2}=\frac{3}{4}$ for all $n$. Furthermore, note that
\begin{equation}
	2\mathbf{S}^{(in)} \cdot \mathbf{S}^{(n')} = \left(\mathbf{S}^{(n)}+\mathbf{S}^{(n')}\right)^2-\mathbf{S}^{(n)2}-\mathbf{S}^{(n')2}=\left(\mathbf{S}^{(n)}+\mathbf{S}^{(n')}\right)^2-\frac{3}{2}.
\end{equation}
Since $\left|\psi\right\rangle$ is symmetric under the exchange of any two spins, any pair of spins add to a total spin of $1$, and thus 
\begin{equation}
\left(\mathbf{S}^{(n)}+\mathbf{S}^{(n)}\right)^2\left|\psi\right\rangle
=1\left(1+1\right)\left|\psi\right\rangle
=2\left|\psi\right\rangle, \quad\forall n \neq n' .
\end{equation}
Since the number of spins is $2\ell$ and the number of pairs is $\ell\left(2\ell-1\right)$, we get eventually that
\begin{equation}
\mathbf{L}^2 \left|\psi\right\rangle
=\left(
2\ell\times\frac{3}{4} + \ell\left(2\ell-1\right) \times \left(2-\frac{3}{2}\right)
\right)\left|\psi\right\rangle
= \ell\left(\ell+1\right)\left|\psi\right\rangle,
\end{equation}
as desired, which completes the proof of statement 1.

\emph{Statement 2: in the same physical system of $2\ell$ spins, we order the spins. If we have full symmetry under the exchange of spins $n$ and $n+1$, we have full symmetry under the exchange of any two spins.}
This statement is very easy to prove: suppose we wish to exchange the spins $n$ and $n+R$, where $R>1$. We can compose this out of several exchanges of nearest neighbors: first exchange $n,n+1$ with $X_{n,n+1}$. Then, $n+1$ and $n+2$ with $X_{n+1,n+2}$ ans so on, until $X_{n+R-1,n+R}$. Then we go backwards, exchange $n+R-1$ with $n+R-2$, all the way back to $n,n+1$. We get that
\begin{equation}
	X_{n,n+R}=X_{n,n+1}\cdots{X_{n+R-1,n+R}}\cdots X_{n,n+1}.
\end{equation}
Thus, if our state $\left|\psi\right\rangle$ is invariant under all the exchanges of nearest neighboring spins, it is invariant under any exchange of any two spins.

From this we can conclude the statement used througout this work - if for any $i=1,...,2\ell-1$ we have
\begin{equation}
\left(\mathbf{S}^{(n)}+\mathbf{S}^{(n+1)}\right)^2\left|\psi\right\rangle,
\end{equation}
we immediately get that $\mathbf{L}^2\left|\psi\right\rangle=\ell\left(\ell+1\right)\left|\psi\right\rangle$, that is, we are in the highest spin multiplet.

\bibliography{ref}

\end{document}